\documentclass[aps,prl,preprint,superscriptaddress]{revtex4-2}
\usepackage{graphicx}
\usepackage{dcolumn}
\usepackage{bm}
\usepackage{xcolor}
\usepackage{xeCJK}
\usepackage{fontspec}
\usepackage{xeCJK}
\usepackage{fontspec}
\usepackage{bibunits}
\defaultbibliographystyle{apsrev4-2}
\makeatletter
\providecommand{\href@noop}[2]{#2}
\makeatother
\begin{document}
\begin{bibunit}
	\preprint{}
	
	\title{Scale-dependent force balance governs transition to the geostrophic regime in liquid metal rotating convection}
	\author{Shao-Peng Yang}
	\affiliation{State Key Laboratory for Strength and Vibration of Mechanical Structures and School of Aerospace, Xi'an Jiaotong University, Xi'an 710049, China}
	\affiliation{Center for Complex Flows and Soft Matter Research and Department of Mechanics and Aerospace Engineering,Southern University of Science and Technology, Shenzhen 518055, China}
	\author{Lin Sun}
	\affiliation{Center for Complex Flows and Soft Matter Research and Department of Mechanics and Aerospace Engineering, Southern University of Science and Technology, Shenzhen 518055, China}
	\author{Guang-Yu Ding}
	\affiliation{Research Institute of Intelligent Complex Systems, Fudan University, Shanghai, China}
	\affiliation{Center for Complex Flows and Soft Matter Research and Department of Mechanics and Aerospace Engineering, Southern University of Science and Technology, Shenzhen 518055, China}
	\author{Ke-Qing Xia}
	\affiliation{Center for Complex Flows and Soft Matter Research and Department of Mechanics and Aerospace Engineering, Southern University of Science and Technology, Shenzhen 518055, China}
	\affiliation{Department of Physics, Southern University of Science and Technology, Shenzhen 518055, China}
	\author{Yi-Chao Xie}
	\affiliation{State Key Laboratory for Strength and Vibration of Mechanical Structures and School of Aerospace, Xi'an Jiaotong University, Xi'an 710049, China}

	\date{\today}
	
	\begin{abstract}
	Rotating convection in low-Prandtl-number liquid metal drives dynamo action in the Earth's outer core and is central to planetary interior dynamics. It has been proposed that flow regime transitions in rotating convection are controlled by competition between the thermal and Ekman boundary layers. However, through laboratory experiments and direct numerical simulations of rotating liquid-metal convection, we find that this mechanism breaks down in the low-Prandtl-number regime. Here we show that increasing rotation reorganises the bulk flow: the large-scale circulation is suppressed and replaced by smaller-scale structures, producing a characteristic horizontal length scale $\ell$.  Transitions to the geostrophic regime are then governed by a buoyancy–Coriolis balance defined on $\ell$ rather than by the boundary-layer crossing. This scale-dependent mechanism also yields heat-transport scalings that depart from boundary-layer-based predictions in the geostrophic regime. Our results reveal a distinct route to the geostrophic regime in low-Prandtl-number rotating convection with implications for rotating liquid metal flows in planetary interiors.
			
	\end{abstract}

	\maketitle

	Thermal convection subject to rotation is a fundamental process in planetary interiors and other geophysical and astrophysical systems. In particular, the geomagnetic field is thought to be generated by rotating convection in the Earth's liquid metal outer core \cite{Stanley2009SSR,Roberts2013RPG,Aurnou2015PEPI,Jones2015BOOK}. The parameter regime relevant to such systems is extreme with estimates for the Earth's core reaching Ekman number (characterising the effects of rotation) as low as $Ek\sim10^{-15}$, Rayleigh number (strength of buoyancy) up to $Ra\sim10^{25}$ and Prandtl number (the ratio of momentum diffsion to heat diffusion) as $Pr\sim10^{-2}$ \cite{Gubbins2001PEPI}. These conditions remain far beyond the reach of present-day laboratory experiments and direct numerical simulations, so the boundaries between different flow regimes, and thus the flow state in celestial bodies must be inferred by extrapolating results obtained under accessible laboratory conditions \cite{Kunnen2021JT}. 
	
	The majority of existing studies have focused on moderate $Pr$ number fluids ($Pr\sim1$) \cite{Rossby1969JFM,King2009Nature,Zhong2009PRL,King2012JFM,Ecke2014PRL,Lu2021PRF,Sun2024PRF}, whereas convection in planetary interiors occurs in liquid metals with $Pr\ll1$, for which thermal diffusion greatly exceeds momentum diffusion. A widely accepted picture of rotating convection based on fluids with $Pr \sim 1 $ assumes that flow regime transitions are controlled by boundary-layer dynamics, in particular by the crossover between the thermal boundary layer and the Ekman boundary layer \cite{King2009Nature}. Based on this idea, a scaling description of the heat transport in rotating convection is developed \cite{Ecke2023ARFM}. 
		
However, rotating convection in the low-$Pr$ number regime is known to differ markedly from its moderate-$Pr$ counterpart. For example, it exhibits oscillatory onset \cite{Chandrasekhar1961book,Rossby1969JFM,Zhang2009JFM} and a strong sensitivity of heat transport to the internal flow structure \cite{Aurnou2018JFM,Fan2024JFM}. These differences raise a central question: does the transition to the geostrophic regime in low-$Pr$ rotating convection follow the same boundary-layer-based picture established in moderate-$Pr$ fluids? Here, we address this problem using a combination of laboratory experiments and direct numerical simulations of liquid metal rotating convection with $Pr=0.029$. We provide evidence that the regime transitions in rotating liquid metal convection are governed by a scale-dependent force balance mechanism rather than the boundary layer crossing mechanism.

\subsection{The flow regime transitions}	
	
		\begin{figure}
    	\includegraphics[width=0.95\linewidth]{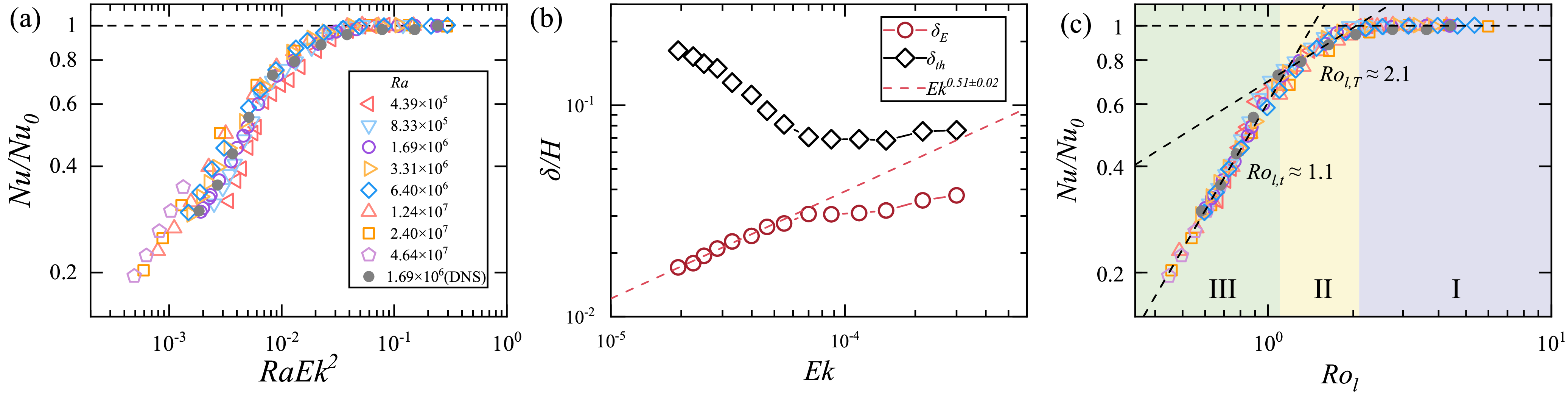}
	    \caption{\textbf{Mechanism of flow-regime transitions in liquid metal rotating convection with $Pr=0.029$.} (a) The normalized heat transport $Nu/Nu_0$ as a function of $RaEk^2$. The choice of x-axis is derived from the boundary layer crossing mechanism \cite{Ecke2023ARFM}. In the rapidly rotating regime with $RaEk<10^{-2}$, the data for different $Ra$ behave differently. (b) The Ekman boundary layer thickness $\delta_E/H$ and the thermal boundary layer thickness $\delta_{th}/H$ as functions of $Ek$ from DNS at $Ra=1.69\times10^6$. The dashed line is a power law fit to the data with $\delta_E/H \sim Ek^{0.51\pm0.02}$. No crossing of the two boundary layers is observed over the explored transition range. The data in (a,b) suggest that the boundary layer crossing mechanism is not applicable in liquid metal rotating convection. (c) The same $Nu/Nu_0$ data plotted against the scale-dependent Rossby number $Ro_\ell$, which characterises the relative importance of buoyancy and Coriolis forces at the scale $\ell$. The data collapse onto a single curve with distinct transitions at $Ro_{\ell,T}\approx2.1$ and $Ro_{\ell,t}\approx1.1$, separating the rotation-unaffected regime I, rotation-influenced regime II, and geostrophic regime III.}
	    \label{fig1}
	 \end{figure}

	We begin by examining the flow regime transitions using the normalised heat transport $Nu/Nu_0$. Here $Nu_0=0.19Ra^{0.25}$ is the heat transport scaling in liquid metal convection without rotation \cite{Ren2022JFM}. Figure \ref{fig1} summarises the central observations.  If the flow regime transition were controlled by the boundary-layer crossing mechanism, $Nu/Nu_0$ would collapse when plotted against $RaEk^2$ \cite{Ecke2023ARFM}. However, as shown in Fig. \ref{fig1}a, the normalised heat transport $Nu/Nu_0$ fails to collapse when plotted against $RaEk^2$ with systematic deviations emerging in the rapidly rotating regime with $RaEk<10^{-2}$. This breakdown is not a quantitative discrepancy but a qualitative failure: data for different Rayleigh numbers behave differently instead of approaching a universal curve. 
	
	To directly assess the underlying assumption of the boundary-layer mechanism, we compare the thermal boundary layer thickness $\delta_T/H$ and the Ekman boundary layer thickness ​$\delta_E/H$. As shown in Fig. \ref{fig1}b, the two boundary layers remain well separated over the $Ek$ range explored with no evidence of a crossover. This observation clearly rules out the boundary-layer crossing mechanism as the origin of regime transitions in the present system.

	We therefore seek the controlling mechanism in the bulk flow. As shown in Fig. \ref{fig1}c, when the same $Nu/Nu_0$ data are plotted against a scale-dependent Rossby number $Ro_\ell$ to be defined below, which charaterises the ratio of boyancy to Coriolis force over a scale $\ell$, the data collapse onto a single master curve with well-defined transitions: $Ro_{l,T}\approx 2.1$ marking the departure from a rotation-unaffected regime I to a rotation-influenced regime II, and $Ro_{l,t}\approx 1.1$ marking the transition to the geostrophic regime III. These results indicate that the flow regime transitions in the low-Pr-number limit are not controlled by the boundary-layer crossing mechanism, but by a scale-dependent buoyancy–Coriolis force balance in the bulk.

 \begin{figure}[htbp]
		\includegraphics[width=0.95\linewidth]{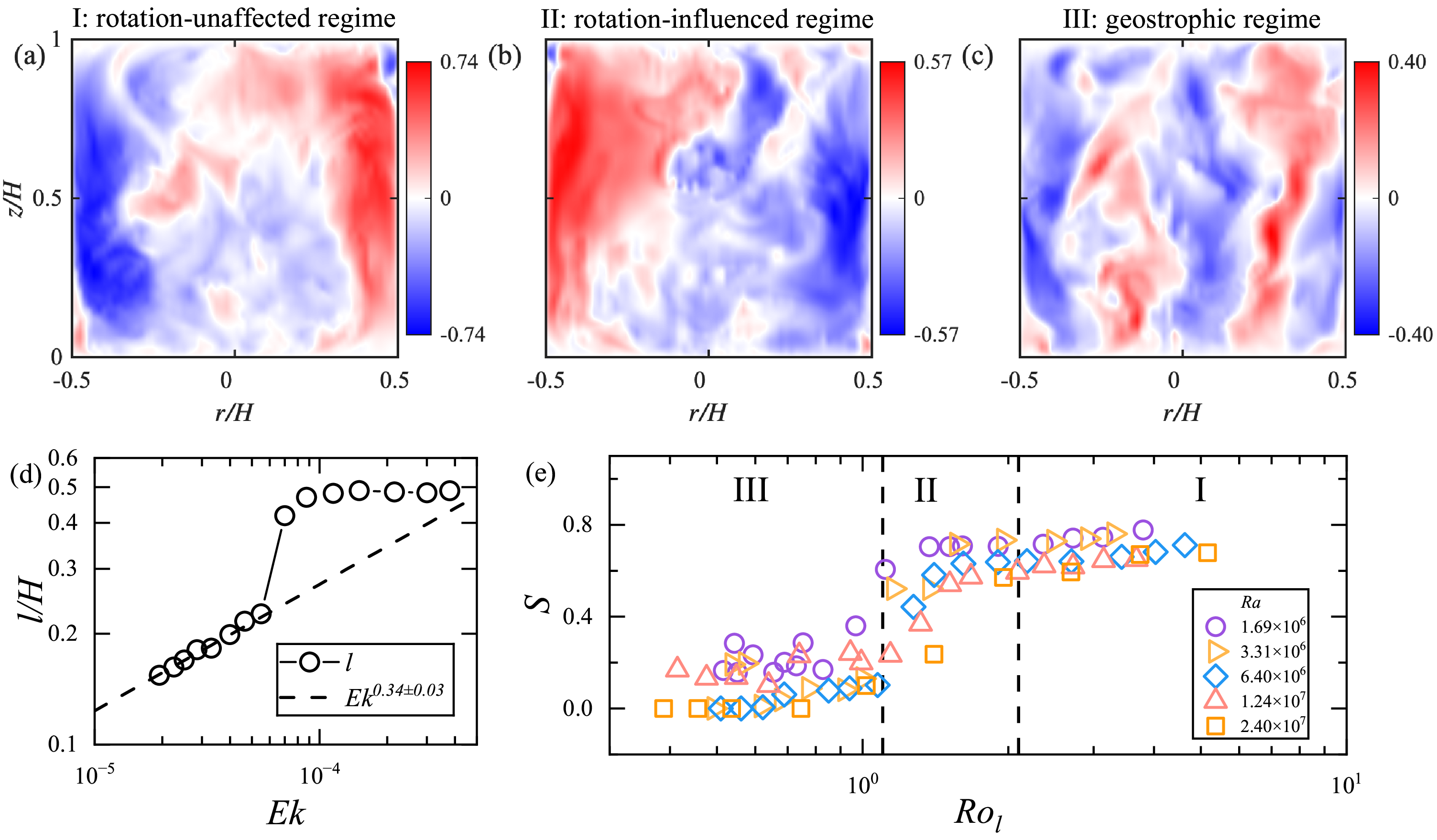}
	\caption{\textbf{Flow organisation and the characteristic length scale.} (a--c) Snapshots of the nondimensionalized vertical velocity in a vertical plane through the cell centre obtained from DNS at $Ra=1.69\times10^6$. The results show the three flow regimes:  the rotation-unaffected regime I with $Ek=3.79\times10^{-4}$, the rotation-influenced regime II with $Ek=7.00\times10^{-5}$, and the geostrophic regime III with $Ek=5.50\times10^{-5}$. With increasing rotation rate, the system-sized large-scale circulation is suppressed and replaced by vertically coherent and laterally confined structures. (d) Characteristic horizontal scale $\ell/H$ as a function of $Ek$. Here $\ell$ is obtained from the zero-crossing of the spatial autocorrelation of the vertical velocity. In regimes I and II where LSC exists, $\ell$ is approximately 0.5. With increasing rotation rate the flow structures change, resulting in sharp decrease of $\ell$. The dashed line is a power law fit to the data in rapidly rotating regime with $\ell/H=6.43Ek^{0.34\pm0.03}$. (e) Relative large-scale circulation strength $S$ measured experimentally as a function of $Ro_\ell$ for different $Ra$. The vertical dashed lines mark the transitions between different flow regimes obtained from the heat transfer measurements.}
	\label{fig2}
	\end{figure}

\subsection{The flow organisation and length scales}

To identify the physical origin of the transition behaviour observed in Fig. \ref{fig1}c, we examine how rotation reorganises the flow in the bulk. Figure \ref{fig2}(a–c) shows vertical velocity fields obtained from DNS at a fixed Rayleigh number $Ra=1.69\times10^6$ across the three regimes. In the rotation-unaffected regime I, the flow is dominated by a system-sized large-scale circulation (LSC) \cite{Ahlers2009RMP,Zhong2010JFM}. As rotation increases, the LSC weakens in regime II and eventually breaks down, giving way to vertically coherent and laterally confined structures in the geostrophic regime III \cite{Kunnen2008EPL}. This transition reflects a fundamental reorganisation of the flow from a global mode to smaller-scale structures set by rotation. Such a reorganisation of the flow structure introduces an emergent horizontal length scale that is absent in the non-rotating system.

 To quantify this scale, we analyse the spatial autocorrelation function of the vertical velocity $R_{ww}$ measured at the mid-height horizontal plane of the convction cell, and we define the  length scale $\ell$ as its first zero-crossing point \cite{Madonia2021EPL,Nieves2014POF}.  As shown in Fig. \ref{fig2}(d), In the weakly rotating regimes I and II where a persistant LSC exists, the measured length scale $\ell$ normalized by the cell height $H$ is approximately 0.5. As rotation increases, the LSC breaks down abruptly and $\ell$ decreases sharply, following a clear power-law dependence on the Ekman number with $\ell/H \approx 6.43Ek^{0.34\pm0.03}$. The observed scaling in rapaidly rotating regime is consistent with the horizontal length scale predicted by linear stability theory for oscillatory convection onset with $\ell/H \approx 2.43\,(Ek/Pr)^{1/3}$ \cite{Chandrasekhar1961book}, suggesting that the observed structures inherit their characteristic size from the onset of convection \cite{Julien2012GAFD,King2013JFM,Aurnou2018JFM}.

The emergence of the rotation-selected length scale $\ell$ provides a natural basis for defining the scale-dependent Rossby number introduced in Fig. \ref{fig1}c. Using the measured $\ell$, we define
        \begin{equation}
	Ro_\ell \equiv \sqrt{ \frac{\alpha g \Delta T}{4\Omega^2 \ell}}
	= \sqrt{\frac{Ra\,Ek^{1.66}}{6.43\,Pr}},
	\label{equ2}
	\end{equation}
which characterises the relative importance of buoyancy and Coriolis forces evaluated at the scale $\ell$ determined from the flow structure. The observation that the transition to the geostrophic regime where rotation becomes dominant occurs at $Ro_{\ell}\approx 1$ (Fig. \ref{fig1}c) therefore indicates that this transition is associated with a balance between buoyancy and Coriolis forces acting on this emergent scale.

This interpretation is further supported by independent flow diagnostics. As shown in Fig. \ref{fig2}(e), the experimentally measured LSC strength $S$ as defined in \cite{Kunnen2011JFM} exhibits clear changes across the same $Ro_{\ell}$​ range that marks the regime transitions in heat transport. 

Taken together, these results show that rotation reorganises the bulk flow and selects a characteristic horizontal scale $\ell$. The consistency between heat transport and flow structures indicates that the flow regime transitions in liquid metal rotating convection are associated with a balance between buoyancy and Coriolis forces acting on this emergent scale.

	\begin{figure}[h]
		\includegraphics[width=\linewidth]{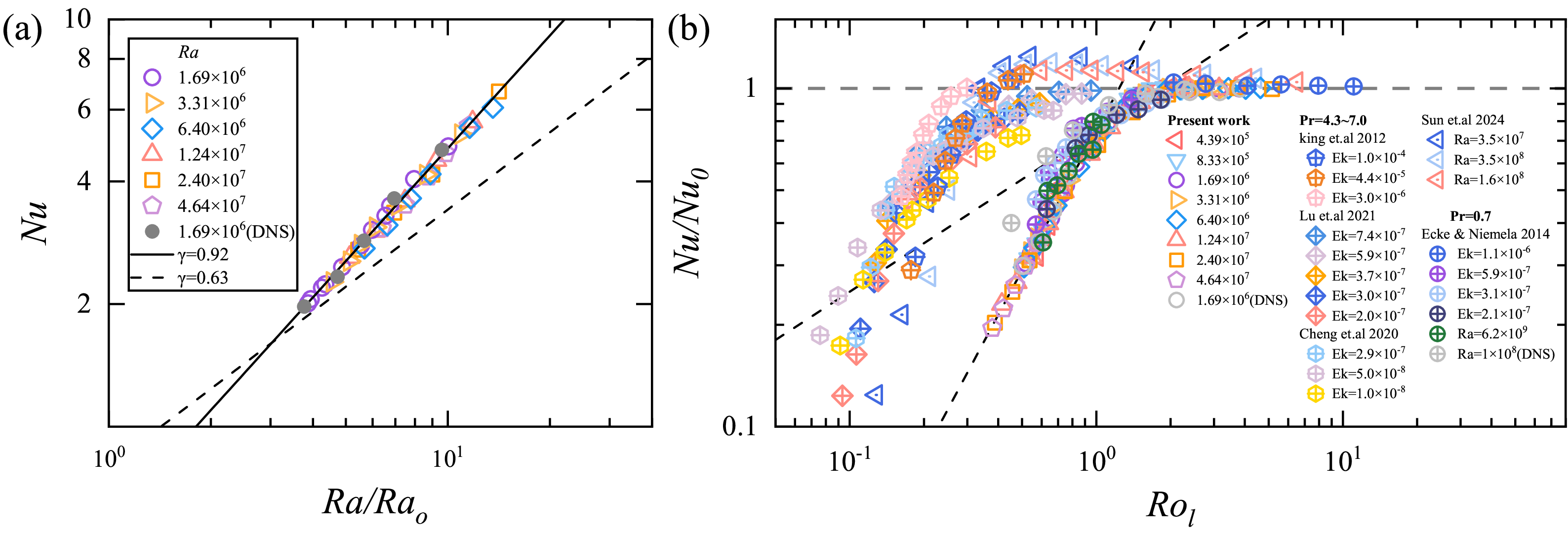}
		\caption{\textbf{{The effects of $Pr$ in the geostrophic regime.}} (a) Nusselt number $Nu$ in geostrophic regime as a function of the supercriticality $Ra/Ra_o$. The open symbols denote experimental data, and the filled symbols represent DNS data. The solid line shows a power law fit to the data with $Nu=(0.58\pm0.02)(Ra/Ra_o)^{0.92\pm0.01}$. The exponent 0.92 is consistent with the prediction based on the transition constraint $Ro_{\ell} \sim 1$. The dashed line shows the scaling relation $Nu\sim(Ra/Ra_o)^{0.63}$ predicted by the boundary-layer-crossing mechanism. (b) Normalised heat transport $Nu/Nu_0$ as a function of $Ro_\ell$ for the present liquid metal rotating convection with $Pr=0.029$, together with data published in the literature for air with $Pr=0.7$ \cite{Ecke2014PRL} and water with $Pr=4.3$--$7.0$ \cite{King2012JFM,Lu2021PRF,Cheng2020PRF,Sun2024PRF}. For this cross-$Pr$ comparison, $Ro_\ell$ is evaluated using the onset-selected horizontal scale from linear stability theory \cite{Chandrasekhar1961book}. The air data follow the same branch as the present low-$Pr$ data and the transition to the geostrophic regime occur near $Ro_\ell\approx1$. However, the water data lie on a distinct branch. These observations suggest a crossover between the scale-dependent force balance mechanism and the boundary layer crossing mechanism around $Pr \approx 0.7$.}
		\label{fig3}
	\end{figure}

\subsection{Scaling consequences of the transition mechanism}
	
	Having identified a scale-dependent transition criterion, we now examine its consequence for heat transport in the geostrophic regime. In non-rotating liquid-metal convection, heat transport follows $Nu_0\sim Ra^{0.25}$ \cite{Ren2022JFM}. In the geostrophic regime, by contrast, the natural scaling variable is the supercriticality relative to the oscillatory onset, so that $Nu\sim (Ra/Ra_o)^{\gamma}$, where $Ra_o\sim Ek^{-1.21}$ for a cylindrical cell with $\Gamma=1$ obtained from \cite{Horn2017JFM,Zhang2009JFM} is used in the present study .
	
	The transition mechanism identified above constrains how the rotation-unaffected regime and the geostrophic regime are connected. At the transition to the geostrophic regime, the heat transport must remain consistent with the non-rotating scaling, while the transition itself is set by the condition $Ro_{\ell}\sim 1$, which yields $Ra_t\sim Ek^{-1.66}$ from equation \ref{equ2}. Combining these relations yields
\begin{equation}
	Nu_t\sim Ra_t^\alpha \sim (Ra_t/Ra_o)^{\gamma}
	\end{equation}
from which one obtains $ \gamma\approx 3.69\alpha$. For $\alpha=0.25$ obtained in liquid metal without rotation \cite{Ren2022JFM}, this predicts $Nu\sim (Ra/Ra_o)^{0.92}$ in the geostrophic regime.

Figure \ref{fig3}(a) tests this prediction. The experimental and DNS data collapse well when plotted against $Ra/Ra_o$, and can be described by
\begin{equation}
Nu=(0.58\pm0.02)(Ra/Ra_o)^{0.92\pm0.01}.
\end{equation}
The measured exponent 0.92 is found to be in excellent agreement with the predicted one. We note this agreement in scaling exponent is not a fitting result, but follows directly from the transition constraint $Ro_{\ell}\sim 1$. However, the conventional boundary-layer-crossing mechanism predicts a much smaller exponent, $\gamma=0.63$, which clearly can not describe the measured data (see Fig \ref{fig3}(a)). The heat-transport scaling in the geostrophic regime therefore carries a clear signature of the underlying transition mechanism.

\subsection{Dependence on Prandtl number}

We next examine how the transition mechanism varies with Prandtl number. To compare fluids with different $Pr$, we evaluate $Ro_{\ell}$  using the onset length scale from linear stability theory \cite{Chandrasekhar1961book}, which preserves the transition behaviour obtained using the measured $\ell$ in Fig. \ref{fig1}c (See also supplementary material for a comparison of the two choices of $\ell$ on the flow regime transitions). Figure \ref{fig3}(b) shows $Nu/Nu_0$ versus $Ro_\ell$ for two groups of data, one with $Pr<1$ from the present study and \cite{Ecke2014PRL}, and the other with $Pr>1$ from \cite{King2012JFM,Lu2021PRF,Cheng2020PRF,Sun2024PRF}.
The data of $Pr<1$ collapses onto a common branch with the transition to geostrophic regime occurring near $Ro_{\ell}\approx 1$. In contrast, the scale-dependent $Ro_{\ell}$ can not bring the data of $Pr>1$ to collapse onto a single curve. In addition, the two $Pr$ groups of data show a large deviation. These observations suggest that the scale-dependent force balance remains relevant up to $Pr\sim 1$, while boundary-layer-based control becomes dominant at higher Prandtl numbers \cite{King2009Nature,Ecke2023ARFM}. Rather than marking a sharp threshold, the present comparison points to a crossover between the two transition mechanisms with $Pr\approx 0.7$ serving only as an empirical indicator within the available data range.

\subsection{Discussion}

The marked difference between flow regime transition mechanisms across $Pr$ (Fig. \ref{fig3}(a)) has important consequences for astrophysical and planetary liquid core dynamics. As noted in previous studies, even modest differences in transition scaling exponents can lead to orders-of-magnitude discrepancies in heat transport when extrapolated to the extreme parameter regimes relevant to geophysical and astrophysical systems \cite{Christensen2009SSR,King2009Nature,King2013PNAS}.

The present results provide a different physical picture of regime transitions in rotating convection at low $Pr$ number. Rather than being controlled by boundary-layer crossing, the transition to the geostrophic regime is associated with a balance between buoyancy and Coriolis forces acting on a rotation-selected bulk length scale. This shift from boundary-layer-controlled to bulk-controlled dynamics fundamentally alters both the transition criterion and the resulting heat-transport scaling.

We now consider the implications of our findings for planetary interior dynamics. Both the Earth and Jupiter host dynamos driven by rapidly rotating, low $Pr$, electrically conducting fluids, yet they differ greatly in their control parameters with $Ek\sim10^{-15}$ and $Pr\sim10^{-2}$ for the Earth \cite{King2010GGG} and $Ek\sim10^{-19}$ and $Pr\sim10^{-1}$ for Jupiter \cite{King2013PNAS}. Despite these differences, their magnetic fields exhibit striking similarities, including dominant axial dipoles, comparable dipole tilting angles (Earth $\sim10^\circ$, Jupiter $\sim11^\circ$), and similar ratios  (both $\sim0.61$)  between dipole and higher-order multipole components \cite{Christensen2006GJI}. These observations suggest a common dynamical origin, but the fluid-mechanical constraints underlying this similarity have remained unclear.
 
 To estimate the flow states of planetary interior within the present framework, it is more appropriate to use the flux Rayleigh number $Ra_F=RaNu$, which is better constrained than $Ra$ in planetary settings. Using the transition criterion $Ro_{\ell} \approx 1$ to the geostrophic regime, which gives $Ra_t\approx 2.41Ek^{-5/3}Pr^{2/3}$, and combining it with the non-rotating heat-transport scaling at the transition $Nu_t=\frac{1}{2} (Ra_tPr)^{1/4}$  \cite{Jones1976JFM,King2013PNAS}, we obtain a transitional flux Rayleigh number
\begin{equation}
	Ra_{F,t}\approx 1.5 Ek^{-25/12}Pr^{13/12}.
\end{equation}

For the Earth with reported values $Ra_F \approx 2.4\sim24\times10^{28}$ \cite{King2010GGG}, we estimate $Ra_F/Ra_{F,t}\approx 0.13\sim1.3$. Similarly, for Jupiter with Ra$_F\approx4\times10^{37}$ \cite{King2013PNAS}, we obtain $Ra_F/Ra_{F,t} \approx 0.1$. While these estimates are necessarily approximate due to uncertainties in planetary parameters and scaling extrapolations, they consistently place both systems near the transition to the geostrophic regime.

This result suggests that Earth and Jupiter may operate in similar dynamical states, close to the boundary between rotation-influenced and geostrophic regimes. We therefore propose that the observed similarities in their magnetic field morphology arise from a shared flow regime constrained by the scale-dependent force balance identified here. More broadly, our results indicate that planetary dynamos may be organised around regime transitions governed by bulk-scale force balances, providing a new perspective for interpreting convection and magnetic-field generation in planetary interiors.

\makeatletter
\putbib[YSDXX2026.bib]
\makeatother
\end{bibunit}
\clearpage

\begin{bibunit}
\section{Methods}
\renewcommand{\figurename}{Extended Data Fig}
\setcounter{figure}{0}

\subsection{Experimental and numerical set-ups}
\begin{figure}[h!]
	\includegraphics[width=0.95\linewidth]{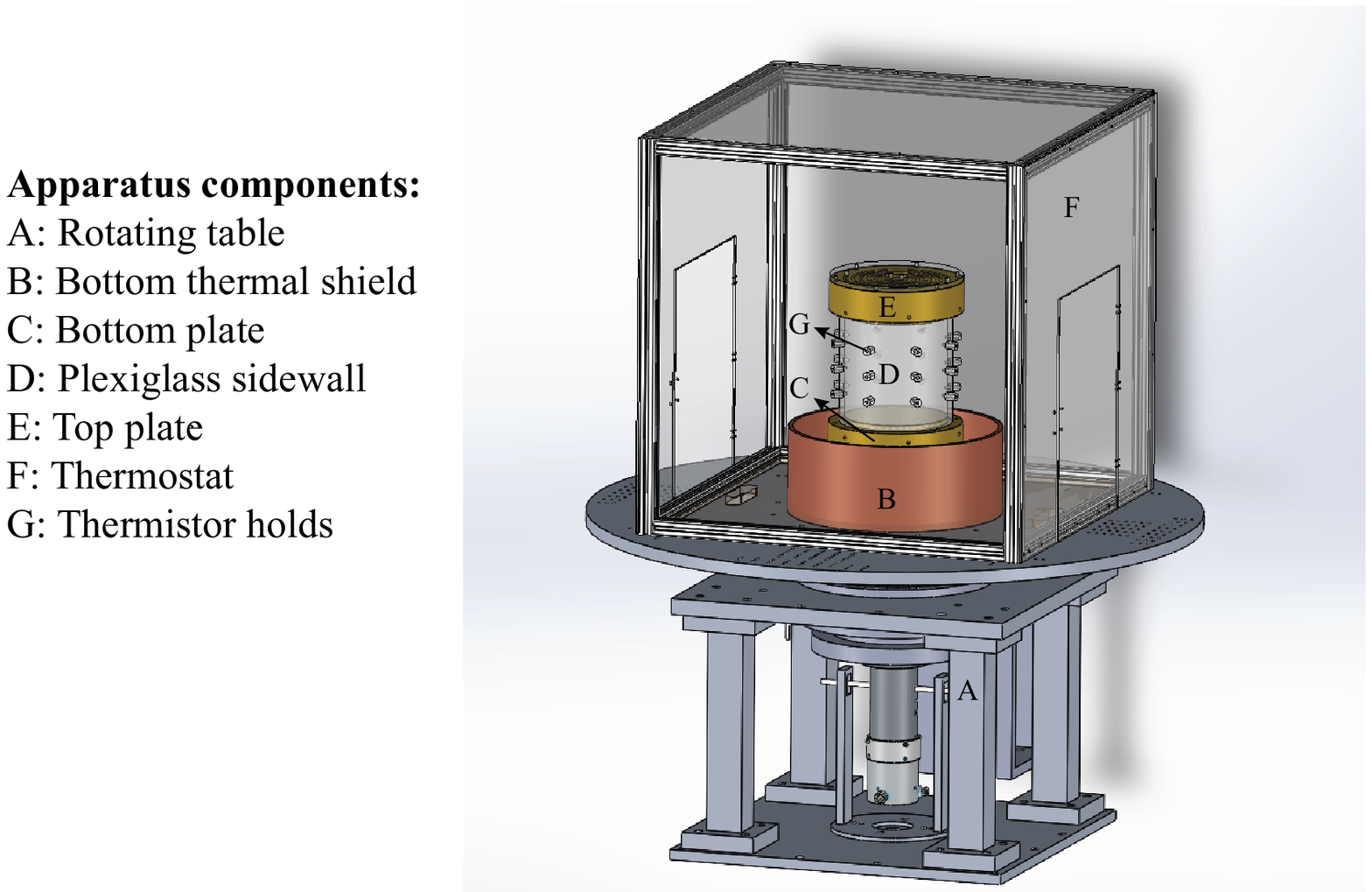}
	\caption{
		Schematic of the experimental setup. A cylindrical convection cell containing liquid-metal alloy gallium-indium-tin (GaInSn) is mounted at the center of the rotating plate and surrounded by a thermostat. The CAD rendering is intended to illustrate the main components and their arrangement and is not drawn to scale.
	}
	\label{fig.setup}
\end{figure}

Laboratory experiments of rotating liquid metal convection were carried out in two upright cylindrical cells with aspect ratio $\Gamma \approx 1$. The two cells had heights $H=99.69~\mathrm{mm}$ and $186.00~\mathrm{mm}$, and diameters $D=101.94~\mathrm{mm}$ and $190.49~\mathrm{mm}$, respectively. The working fluid is the liquid-metal alloy gallium-indium-tin (GaInSn). Its mean temperature was fixed at $35\,^{\circ}\mathrm{C}$, at which the Prandtl number is $Pr=0.029$. 

All rotating experiments were performed on a home-built rotating table (A) as shown in Extended Data Fig \ref{fig.setup}. The overall design of the rotating table has been described in detail \cite{Hu2022JFM,Hu2025RSI}. Heat was supplied from below by an electrical resistive heater attached to the bottom plate (C) and removed from above by a thermostated recirculating water bath connected to the top plate (E), thus the bottom plate is at the constant heat flux boundary condition and the top plate is at the constant temperature boundary condition. The temperatures of the top and bottom plates were measured by four thermistors embedded into the plate. The imposed temperature difference $\Delta T$ across the convection cell was determined from the mean plate temperature. To measure the dynamics of the large-scale flow from the sidewall temperature, the muti-thermal probe method was employed \cite{Cioni1997JFM}. Thermistors were inserted into blind holes throgh thermistor holds (G) on the sidewall and were distributed in three horizontal rows with altitude $H/4$, $H/2$, and $3H/4$ from the bottom plate, and in eight vertical columns equally spaced azimuthally. The convection cell was wrapped with rubber foam and placed in the bottom thermal shield (B) to reduce heat loss. The temperature of the shield was adjusted to be equal to that of the bottom plate to minimize heat leakage. In addition the whole cell was placed inside a thermostat (F) to keep it isolated
from the surrounding temperature variation.

A total of 119 experimental runs were carried out spanning a Rayleigh number range of $4.39\times10^5 \le Ra \le 4.64\times10^7$ and Ekman number range of $3.25\times10^{-6}\le Ek \le 4.32\times10^{-4}$. For each experimental run with fixed $Ra$, the target non-rotating $Ra$ was first established. The rotating table was then accelerated slowly to the prescribed angular velocity $\Omega$ ($Ek$), after which the heating and cooling powers were carefully adjusted so that the $Ra$ remained nearly constant for different values of $Ek$. Once the system reached a stationary state, temperature data were recorded for at least 8 hours. The heat flux $q$ was obtained from the net heating power and the Nusselt number was calculated as
\begin{equation}
	Nu=\frac{qH}{\lambda \Delta T},
\end{equation}
where $\lambda=24.9~\mathrm{W\,(mK)^{-1}}$ is the thermal conductivity of GaInSn evaluated at the mean fluid temperature $35\,^{\circ}\mathrm{C}$. The control parameters were determined as
\begin{equation}
	Ra=\frac{\alpha g \Delta T H^3}{\nu\kappa}, \qquad
	Ek=\frac{\nu}{2\Omega H^2}, \qquad
	Pr=\frac{\nu}{\kappa},
\end{equation}
using material properties evaluated at the mean fluid temperature \cite{Ren2022JFM}. 

Direct numerical simulation of the governing Oberbeck-Boussinesq equations was carried out in the cylindrical coordinates using a well-tested DNS code called CUPS, which is a fully parallelized DNS code based on the finite volume method with fourth-order precision \cite{Chong2018JCP}. This code has been widely used to study rotating convection in conventional fluids, such as water \cite{Chong2020SA,Lu2021PRF,Ding2023JFM}. In this study we focus on the system with $Pr=0.029$ and $\Gamma=1$. We conducted a series of simulations at a fixed Rayleigh number $Ra=1.69\times 10^6$ and varied Ekman numbers in the range of $1.94\times 10^{-5}\le Ek\le 3.79\times 10^{-4}$. All statistical quantities are taken over 400 convective free-fall time units after the system had reached a stable state.

\subsection{Measurement of the boundary layer thickness using the DNS data}

\begin{figure}[h]
	\includegraphics[width=\linewidth]{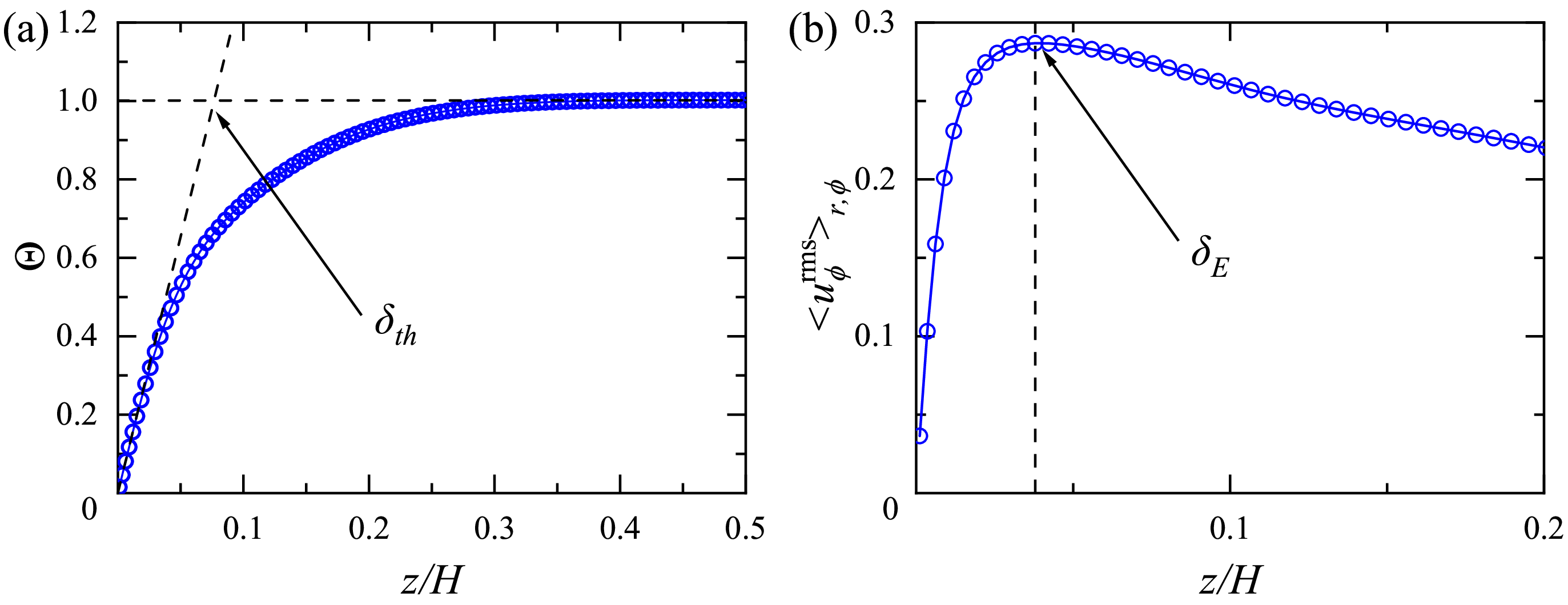}
	\caption{
		Example of boundary-layer thickness measured from statistically averaged profiles for $Ra=1.69\times10^{6}$ and $Ek=3.79\times10^{-4}$. (a) Normalized mean temperature profile $\Theta(z)$.
		The circles show the averaged profile, and the dashed lines show the linear fits used to determine the thermal boundary-layer location. The arrow marks the thermal boundary layer thickness $\delta_{th}$ at the bottom plate, defined from the intersection between the near-wall fit and the bulk fit. (b) Root-mean-square profile of the azimuthal velocity $ \langle u^{\rm rms}_{\phi}\rangle_{r,\phi}$. The vertical dashed line marks the near-wall maximum rms velocity used to define the Ekman boundary layer location. The arrow marks the Ekman boundary-layer thickness $\delta_E$.}
	\label{fig.BL}
\end{figure}

The thermal and velocity boundary-layer thicknesses were measured from the time averaged temperature and root-mean-square velocity fields. As shown in Extended Data Fig \ref{fig.BL} (a), the thermal boundary layer thickness was determined using the so-called slope method using the normalized temperature $\Theta (z)$, defined as 
\begin{equation}
	\Theta(z)=\frac{\theta_{\rm bot}-\langle \overline{\theta}\rangle_{r,\phi}(z)}{\Delta/2}
\end{equation}
where $\theta_{\rm bot}=0.5$ is the temperature of the bottom plate and $\langle \overline{\theta}\rangle_{r,\phi}(z)$ is the mean temperature profile \cite{Zhou2010JFM}. Here angle brackets and overbar represents spatial (radius direction $r$ and azimuthal direction $\phi$) and time averaging, respectively. The Ekman boundary layer thickness was identified as the location where the root-mean-square azimuthal velocity $ \langle u^{\rm rms}_{\phi}\rangle_{r,\phi}$ reaches a maximum as shown in Extended Data Fig \ref{fig.BL} (b).

\subsection{Characteristic length scale $\ell$}
\begin{figure}[htpb]
	\centering
	\includegraphics[width=0.55\linewidth]{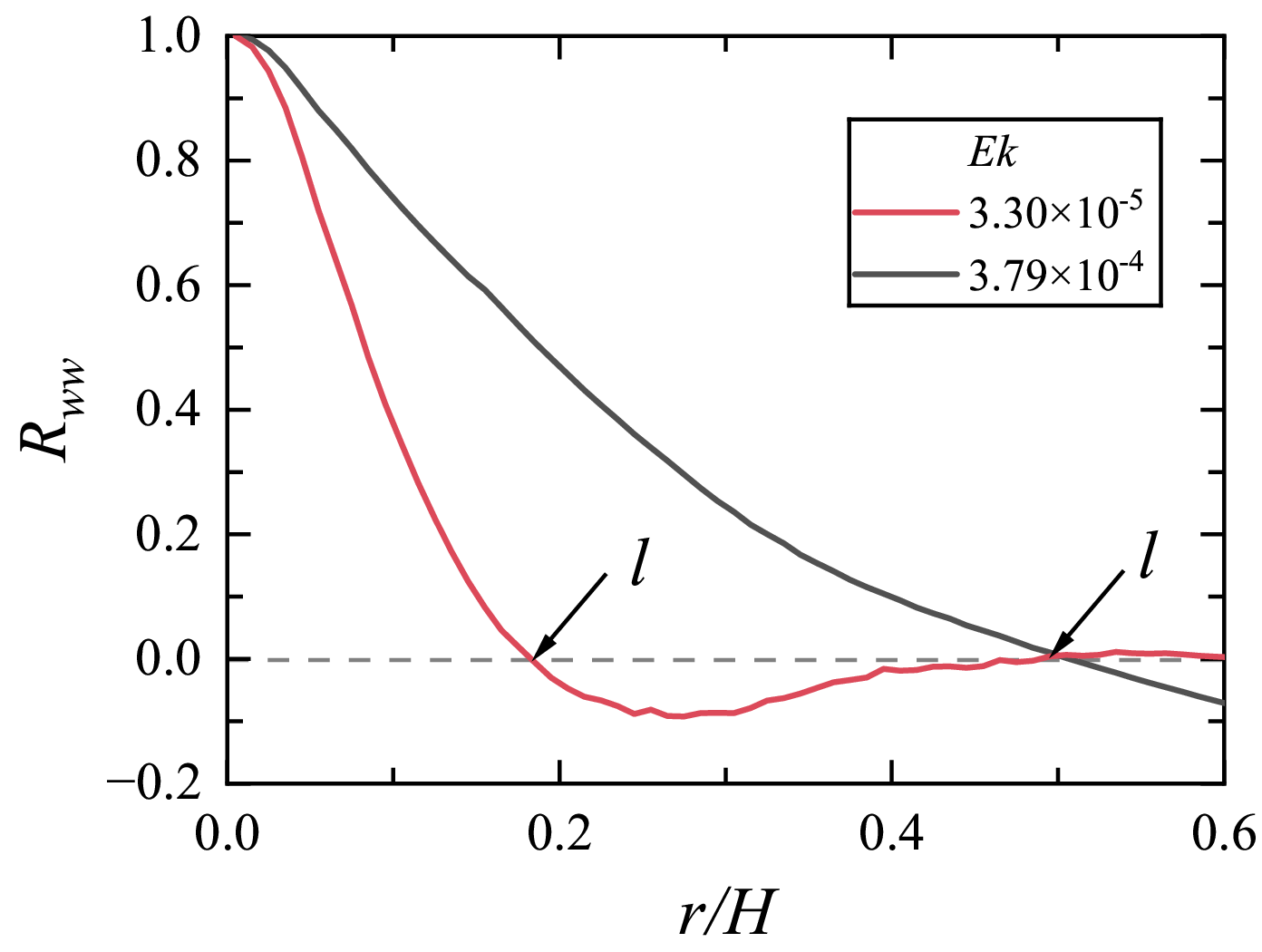}
	\caption{
		Spatial autocorrelation $R_{ww}$ of the vertical velocity $w$ of two cases. The red and black solid line show the data of  $Ek=3.30\times10^{-5}$ in geostrophic regime and $Ek=3.79\times10^{-4}$ in rotation-unaffected regime, respectively. The horizontal length scale determined by the zero crossing point of $R_{ww}$ is denoted as $\ell$.
	}
	\label{fig.Rw}
\end{figure}
To quantify the characteristic horizontal length scale of the flow structures, following previous study of rapidly rotating convection \cite{Madonia2021EPL,Nieves2014POF}, we computed the spatial autocorrelation of the vertical velocity field at the horizontal mid plane. For a scalar field $f({\bf x})$, the normalized spatial autocorrelation is defined as
\begin{equation}
	R_{ff}(r)=
	\frac{\left< f({\bf x}) f({\bf x}+{\bf r}) \right>_{{\bf A}}}
	{\left< f^2({\bf x}) \right>_{{\bf A}}},
\end{equation}
where $r=|{\bf r}|$, and $\left< \cdot \right>_{{\bf A}}$ denotes averaging over all effective area A in the selected horizontal plane. In the present analysis, $f$ is taken as the instantaneous vertical velocity $w$. To focus on the bulk flow, the effective area A for the autocorrelation is reduced to exclude the wall mode \cite{Zhang2020PRL}. The two-dimensional autocorrelation was then calculated using zero-padding and two-dimensional fast Fourier transforms. The resulting two-dimensional correlation field was converted into a one-dimensional radial correlation function by averaging over circular annuli.

The horizontal convective length scale was extracted from the first zero crossing of the radial autocorrelation function as shown in Extended Data Fig \ref{fig.Rw}, which is denoted as $\ell$.

\subsection{Measurement of the relative strength of the large-scale circulation}

The strength of the large-scale circulation (LSC) was quantified from the azimuthal temperature distribution measured by the thermistor arrays embedded in the sidewall \cite{Ahlers2009RMP,Ren2022JFM,Cioni1997JFM,Zürner2019JFM}. To obtain a normalized measure of the LSC strength, following Kunnen et al. \cite{Kunnen2011JFM}, the relative LSC strength is defined as
\begin{equation}
	S=
	\max\left[
	\frac{\overline{E_1}/\overline{E_{\rm tot}}-1/N}{1-1/N},\,0
	\right].
\end{equation}
Here $\overline{E_1}$ indicates the energy in the first Fourier mode which corresponds to the LSC; $\overline{E_{\rm tot}}$ is the sum of the total energy in all Fourier modes; $N$ is the number of resolved Fourier modes and $N=4$ in the present study. $S=1$ indicates a pure cosine azimuthal profile, which is a signature of the LSC and $S=0$ indicates high order modes. So the decrease of S corresponds to the weakening and breakdown of the LSC.

\textbf{Funding}: This work is supported by the National Natural Science Foundation of China under grant Nos. 12595303,12595300,12595301,12422209,12232010.
\makeatletter
\putbib[supplementary.bib]
\makeatother
\end{bibunit}
\end{document}